\documentclass[aps, showpacs, twocolumn]{revtex4}

\usepackage{amsmath,amssymb}
\usepackage{tabularx}
\usepackage{epsfig}
\usepackage{graphicx}

\begin{document}

\title{Nonminimal GUT inflation after Planck results}

\author{Grigoris Panotopoulos\footnote{gpanotop@ing.uchile.cl}}

\date{\today}


\address{Department of Physics, FCFM, Universidad de Chile, Blanco Encalada 2008, Santiago, Chile}


\begin{abstract}
In the present work we study GUT Coleman-Weinberg inflation with a
nonminimal coupling to gravity. In this kind of model one usually
finds that either the nonminimal coupling to gravity is large $\xi
\gg 1$ or the inflaton self-coupling is unnaturally small $\lambda
\sim 10^{-13}$. We have shown that the model is in agreement with
the recent results from Planck for natural values of the couplings.
\end{abstract}


\maketitle

Inflation~\cite{inflation} has become the standard paradigm for the
early Universe, because it solves some outstanding problems present
in the standard hot big-bang cosmology, such as the flatness and
horizon problems, the problem of unwanted relics, such as magnetic
monopoles, and produces the cosmological fluctuations for the
formation of the structure that we observe today. The spectacular
CMB data, first from the WMAP satellite~\cite{wmap} and recently
from  Planck~\cite{planck}, have strengthen the inflationary idea,
since the observations indicate an \emph{almost} scale-free spectrum
of Gaussian adiabatic density fluctuations, just as predicted by
simple models of inflation. However, inflation is not a theory yet,
as we don't know how to integrate it with ideas from particle
physics.

The Brout-Englert-Higgs mechanism~\cite{BEH} in the framework of the
Standard Model breaks electroweak symmetry, gives masses to the
charged fermions and the massive gauge bosons, and predicts the
existence of the Higgs boson. The recent LHC discovery of the Higgs
boson~\cite{LHC} completed the particle spectrum predicted by the
Standard Model, and indicates for the first time that fundamental
scalar particles exist in nature. Therefore, it is a natural thing
to assume that inflation is driven by the Higgs boson. Although the
Higgs potential is not suitable for a viable inflationary model, the
presence of a Higgs non-minimal coupling to gravity can change
things to the better. A large value of the non-minimal coupling $\xi
\sim 10^4$ is required, and given the uncertainties in the top quark
mass and the strong coupling constant the model is still
allowed~\cite{higgsinflation}. Sadly, that large value of $\xi \gg
1$ questions the validity of the scenario in the SM Higgs inflation,
as the inflationary scale exceeds the effective ultraviolet cut-off
scale~\cite{espinosa}.

As it was realized long ago, radiative corrections can be the origin
of electroweak symmetry breaking~\cite{CW}. The Coleman-Weinberg
mechanism cannot work in the Standard Model due to the large value
of the top Yukawa coupling, but it can work in models beyond the
Standard Model~\cite{BSM}. This is an interesting possibility given
the Higgs naturalness problem, in the following sense. If heavy
particles are coupled to the Higgs boson, like in GUT models, then
the Higgs receives large radiative corrections that bring its mass
close to the GUT mass scale. Supersymmetry at the TeV scale can
solve the problem, but given the severe experimental constraints on
the masses of the superpartners, it was proposed recently the
flatland scenario~\cite{flatland}, according to which electroweak
symmetry is broken radiatively a la Coleman-Weinberg in the infrared
region starting from a flat scalar potential in the ultraviolet
region.

Regarding inflation, the Coleman-Weinberg type of potential is a
simple and well motivated one, since it naturally arises when loop
corrections are taken into account, and it is typical for the new
inflation scenario~\cite{new} where inflation takes place near the
maximum. Recently it has been studied in~\cite{gabriela} in a B-L
extension of the Standard Model, and a few years ago in~\cite{shafi}
in a GUT inflationary model. As shown in~\cite{RefD} in the context
of the effective theory of inflation, Cosmic Microwave Background
together with Large Scale Structure data prefer double-well inflaton
potentials. The 2013 Planck data confirmed this result, and not
surprisingly the Coleman-Weinberg potential considered here belongs
to this class, and succeeds to reproduce the $n_s$ value and the r
bound from the 2013 Planck release. However, in~\cite{gabriela,
shafi} the inflaton was minimally coupled to gravity. Setting
$\xi=0$, although it is a popular choice, is often unacceptable as
was pointed out in~\cite{faraoni}. Non-minimal couplings are
generated by quantum corrections even if they are absent in the
classical action~\cite{linde}, and as a matter of fact the coupling
is required if the scalar field theory is to be renormalizable in a
classical gravitational background~\cite{freedman}. For early works
on non-minimal inflation see for example~\cite{old} and references
therein.

In the present Brief Report we imagine a scenario where inflation is
driven by the scalar sector of some particle physics model beyond
the Standard Model in which electroweak symmetry breaking takes
place radiatively a la Coleman and Weinberg. Since the scale of
inflation is close to the GUT scale, we consider a GUT model rather
than a low energy one, and we also allow for a non-vanishing
non-minimal coupling, as required by the quantum corrections already
needed to give rise to the Coleman-Weinberg type of the inflaton
potential. We find that the model is viable as it is in agreement
with the recent data from Planck, and even more importantly for
natural values of the parameters of the model.

We start defining the model by the action
\begin{equation}
S = \int d^4x \sqrt{-g} \left ( \frac{1+\kappa^2 \xi \phi^2}{2
\kappa^2} R-(1/2) \phi_{; \mu} \phi^{; \mu} - V(\phi) \right )
\end{equation}
where $\xi$ is the non-minimal coupling and $\kappa^{-1}=M_p=2.4
\times 10^{18}~GeV$ with a Coleman-Weinberg type of potential of the
standard form~\cite{gabriela,shafi}
\begin{equation}
V(\phi)=A \phi^4 \left ( log(\frac{\phi}{M}) -\frac{1}{4} \right
)+\frac{A}{4} \: M^4
\end{equation}
where A is related to the inflaton quartic
self-coupling~\cite{gabriela,shafi}, and M is the inflaton vacuum
expectation value (vev) at the minimum, $M=M_{GUT} \sim
10^{16}~GeV$. The vacuum energy density at the origin is given by
the constant term $V_0=A M^4/4$ so that $V(\phi=M)=0$, and the shape
of the potential can be seen in Figure~1, and the inflaton mass is
given by $m_{\phi}=2 \sqrt{A} M$. At this point we stress the fact
that the Coleman-Weinberg potential considered in the present work
is the standard one obtained from one-loop corrections in Minkowski
spacetime. The one-loop corrections in de Sitter spacetime (which
approximates an inflationary background) lead to a very different
effective potential as shown in~\cite{RefABC}.

\begin{figure}[h!]
\centering
\begin{tabular}{cc}
\includegraphics*[width=260pt, height=260pt]{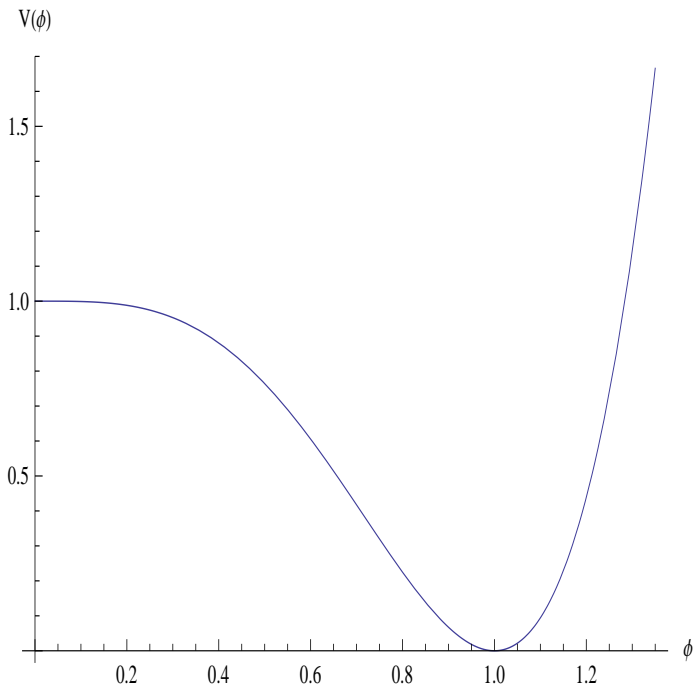}&%
\end{tabular}
\caption{The Coleman-Weinberg potential as a function of the scalar
field.}
\end{figure}

The non-minimal coupling can be eliminated going to the Einstein
frame (in which the scalar field is denoted by $\sigma$ and the
metric by $\hat{g}_{\mu \nu}$) through a conformal transformation
\begin{eqnarray}
\hat{g}_{\mu \nu} & = & \Omega^2 \: g_{\mu \nu} \\
U & = & \frac{V}{\Omega^4} \\
\frac{d \sigma}{d \phi} & = & \frac{\sqrt{1+\kappa^2 \xi \phi^2 (1+6
\xi)}}{\Omega^2}
\end{eqnarray}
where $\Omega^2=1+\kappa^2 \xi \phi^2$.

Assuming that in the Jordan frame inflation takes place near the
maximum of the potential, $\phi < M < M_p$, the potential is
dominated by the constant term $V_0=A M^4/4$, and the model at hand
is a small-field model where the slow-roll parameters are as follows
$\eta < 0, \eta < -\epsilon$~\cite{kinney}. Requiring that $\xi < 1$
we can make the approximation that $\sigma \sim \phi$ and therefore
in the Einstein frame the potential takes the form
\begin{equation}
U(\sigma) \simeq V_0 (1-2 \kappa^2 \xi \sigma^2)
\end{equation}
and it is of the form
\begin{equation}
V(\phi)=\Lambda^4 (1-(\phi/\mu)^2)
\end{equation}
with $\mu=M_p/\sqrt{2 \xi}$. One can easily see that for this model
indeed $\eta < -\epsilon$, and the spectral index and
tensor-to-scalar ratio are given by~\cite{kinney}
\begin{eqnarray}
n_s & = & 1-4 (M_p/\mu)^2 \\
r & = & 8 (1-n_s) exp(-1-N_* (1-n_s))
\end{eqnarray}
and they are independent of $V_0=\Lambda^4$. Here we take the number
of e-foldings to be $N_*=60$, for which the tensor-to-scalar ratio
is computed to be $0.0089 < r < 0.0146$. The model agrees very well
with the recent data from Planck~\cite{planck} for $9 < \mu/M_p <
11$, or
\begin{equation}
0.004 < \xi < 0.006
\end{equation}
At this point we should check the validity of our approximation. We
have neglected the second order term $\kappa^4 \xi^2 \sigma^4$, so
now we need to show that the ratio
\begin{equation}
z=\frac{\kappa^4 \xi^2 \sigma^4}{\kappa^2 \xi \sigma^2}
\end{equation}
is indeed small. Using the definitions it is easy for someone to
show that $r=128 \xi z$, and therefore $z \simeq 0.03$ at most.

Finally the amplitude of the curvature perturbation $\Delta_R=4.9
\times 10^{-5}$ is given by
\begin{equation}
\Delta_R=\frac{U^{3/2}}{2 \sqrt{3} \pi |U|}
\end{equation}
from which we find a relation between the couplings $\xi$ and
$\lambda$, which is the following
\begin{equation}
A(\xi) = \frac{480 \pi^2}{e} \: \left( \frac{M_p}{M} \right)^4 \:
\Delta_R^2 \: \xi^2 exp(-8 N_* \xi)
\end{equation}
and can be seen in Figure~2 for $M=0.01 \: M_p=2.4 \times
10^{16}~GeV$. If we take M to be slightly lower, $M=1.1 \times
10^{16}~GeV$, A becomes of the order $\sim 0.01$, and for $M=6.7
\times 10^{15}~GeV$ the coupling $A \sim 0.1$. Therefore the model
is in agreement with the recent data from Planck, but even more
importantly it manages to be a viable model for natural values of
the couplings $\xi \sim 10^{-3}$ and $A \sim 10^{-4}-10^{-1}$
depending on the precise value of $M=M_{GUT}$. Contrary to the
results found in other
works~\cite{higgsinflation,gabriela,shafi,okada} neither the
non-minimal coupling is large, $\xi \sim 10^4$, nor the inflaton
self-coupling is tiny $A \sim 10^{-14}$. In particular,
in~\cite{higgsinflation} successful inflation requires that the
non-minimal coupling $\xi$ and the SM Higgs quartic self-coupling
$\lambda$ satisfy the relation $\xi \simeq 44700 \: \lambda$. But
since $\lambda$ is determined by the Higgs boson mass, $\lambda
\simeq 0.5$, it turns out that $\xi \sim 10^4$. Similarly, in the
Figure~6 of~\cite{okada} one can see that $\lambda$ increases with
$\xi$, and therefore either $\xi$ is large or $\lambda$ is
unnaturally small. Finally, in~\cite{gabriela, shafi} (where
$\xi=0$) the spectral index requires a transplanckian Higgs vev, $M
\simeq 10M_p$, and then for this value of M the coupling A has to be
of the order of $A \sim 10^{-14}$. In our work, the non-vanishing
non-minimal coupling helps in increasing A keeping at the same time
the Higgs vev at the GUT scale, $M=M_{GUT} \sim 10^{16}$~GeV.

\begin{figure}[h!]
\centering
\begin{tabular}{cc}
\includegraphics*[width=260pt, height=260pt]{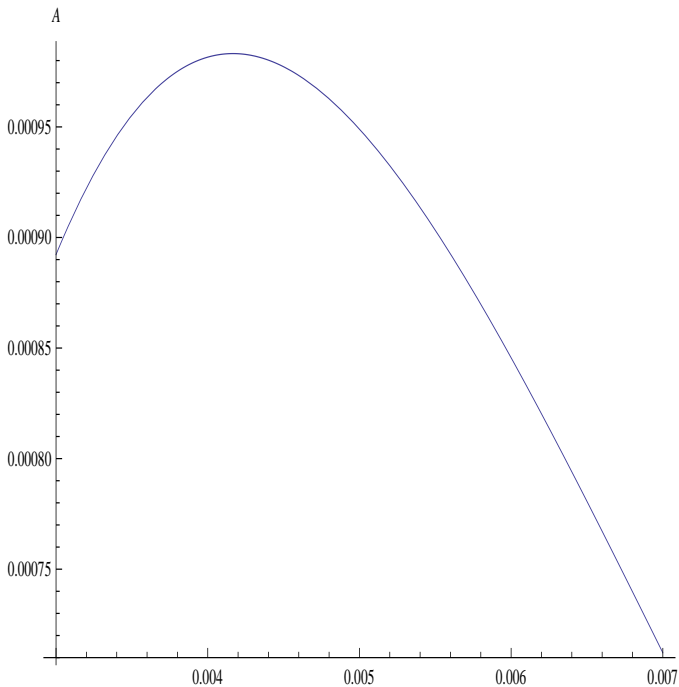}&%
\end{tabular}
\caption{The coupling A as a function of the non-minimal coupling
$\xi$ for $M=0.01~M_p$.}
\end{figure}

In summary, in this Brief Report we have considered a GUT
inflationary model where inflation is driven by the scalar sector of
the model with a Coleman-Weinberg type of potential for the inflaton
and a non-minimal coupling of the inflaton to gravity. As is known,
at tree level the $\lambda \: \phi^4$ potential cannot trigger
electroweak symmetry breaking, and in addition the corresponding
chaotic inflationary model $\phi^4$ is ruled out by the WMAP and
Planck data. However, quantum loop corrections modify the inflaton
potential and at the same time generate the non-minimal coupling
term. The model is characterized by two dimensionless parameters,
namely the non-minimal coupling $\xi$ and the inflaton quartic
self-coupling $\lambda$. We have shown that the model leads to
predictions in agreement with the 95 per cent CL contours in the
r-$n_s$ plane for natural values of the parameters of the model,
both for $\xi$ and $\lambda$ of the order of $10^{-3}$.

\section*{Acknowlegements}

We wish to thank the anonymous reviewer for his valuable comments
that helped in improving the quality of the article. Our work is
supported by Conicyt under the Anillo project ACT1122.

\end{document}